\documentclass[12pt]{iopart}

\usepackage{graphicx}
\usepackage{amssymb}
\usepackage{bbm}
\usepackage{layouts}
\usepackage{color}
    \definecolor{Blue}{rgb}{0.0,0.0,1.0}
    \definecolor{Red}{rgb}{1.0,0.0,0.0}
    \definecolor{Green}{rgb}{0.0,1.0,0.0}

\begin{document}

\title[]{The perihelion of Mercury advance and the light bending calculated
in (enhanced) Newton's theory}

\author{
M.A. Abramowicz$^{1, 2, 3, 4, 5}$                                      %
G.F.R. Ellis$^{4}$                                                     %
J. Hor{\'a}k$^{5, 3}$                                                  %
M. Wielgus$^{6, 3, 5}$                                                 %
}                                                                      %
\address{                                                              %
$^1$Klinika Chirurgii Gastroenterologicznej i Transplantologii         %
               Centralnego Szpitala Klinicznego MSW, Warszawa, Poland\\%
$^2$Physics Department, Gothenburg University,                         %
               SE-412-96 G{\"o}teborg, Sweden \\                       %
$^3$Copernicus Astronomical Center, ul. Bartycka 18,                   %
               PL-00-716, Warszawa, Poland    \\                       %
$^4$Mathematics Department, University of Cape Town,                   %
               Rondebosch, Cape Town 7701, South Africa \\             %
$^5$Astronomical Institute of the Academy of Sciences                  %
               Bo{\v c}ni II 1401/1a,                                  %
               CZ-141 31 Praha 4, Czech Republic   \\                  %
$^6$Institute of Micromechanics and Photonics,                         %
               ul. {\'S}w. A. Boboli 8, PL-02-525,                     %
               Warszawa, Poland                                        %
}
\ead{
marek.abramowicz@physics.gu.se,                                        %
gfrellis@gmail.com,                                                    %
horak@astro.cas.cz,                                                    %
maciek.wielgus@gmail.com                                               %
}
%
\begin{abstract}
We show that results of a simple dynamical {\it gedanken} experiment in\-ter\-pre\-ted
according to standard Newton's gravitational theory, may reveal that three-dimensional space
is curved. The experiment may be used to reconstruct the curved geometry of space, i.e. its
non-Euclidean metric $^3g_{ik}$. The perihelion of Mercury ad\-van\-ce and the light bending
calculated from the Poisson equation $^3g^{ik} \nabla_i \nabla_k \Phi = -4\pi G \rho$ and
the equation of motion $F^i = ma^i$ in the curved geometry $^3g_{ik}$ have the correct
(observed) values. Independently, we also show that Newtonian gravity theory may be enhanced
to incorporate the curvature of three dimensional space by adding an extra equation which
links the Ricci scalar $^3R$ with the density of matter $\rho$. Like in Einstein's general
relativity, matter is the source of curvature. In the spherically symmetric (vacuum) case,
the metric of space $^3g_{ik}$ that follows from this extra equation agrees, to the expected
accuracy, with the metric measured by the Newtonian gedanken experiment mentioned above.
\end{abstract}


\maketitle



\section{Introduction}
%
Newton's theory of gravity was formulated in a flat, Euclidean 3-D space but its basic laws,
i.e. the Poisson equation and the equation of motion,
\begin{eqnarray}
%
%
{}^{3}\,g^{ik}\,\, \nabla_i \nabla_k \Phi = -4\pi G \rho,
\label{Poisson-equation} \\
F_i = m a_i,
\label{equation-of-motion}
\end{eqnarray}
make perfect sense in a 3-D space with an arbitrary geometry ${}^{3}g_{ik}$. Indeed, the
curvature of space is (potentially) present in Newton's theory. It is easy to argue that the
``centrifugal'' acceleration of a particle moving with velocity $V$ on a circular orbit
equals $a_{\scriptscriptstyle C} = V^2/{\cal R}$, and the ``gravitational'' acceleration in
the gravitational field of a spherically symmetric body with the mass ${\cal M}$ equals
$a_{\scriptscriptstyle G} = G{\cal M}/{\tilde r}^2$, with ${\cal R}$ being the curvature
radius of the circle, and ${\tilde r}$ being its circumferential radius. In flat, i.e.
Euclidean, 3-D space these two radii are equal, ${\cal R} = {\tilde r}$, but in a space with
a~non-zero gaussian curvature ${\cal G}$, they are different, ${\cal R} \not = {\tilde r}$.
Therefore, by measuring centrifugal and gravitational accelerations one may independently
measure ${\cal R}$ and ${\tilde r}$, and thus {\it experimentally} find whether the space is
flat (Euclidean) or it has a~non-zero Gaussian curvature ${\cal G} \not = 0$. Based on that,
Abramowicz has recently suggested in \cite{Abramowicz-2012} that a Newtonian physicist could
experimentally determine the metric ${}^{3}g_{ik}$ of the real physical 3-D space and
calculate, according to (\ref{Poisson-equation}) and (\ref{equation-of-motion}), the
perihelion of Mercury advance and the light bending effects. In this paper we follow this
suggestion and calculate both effects within Newton's theory. Surprisingly, the values of
the perihelion advance and the light bending agree (to the expected order of $M/r$) with
predictions of Einstein's theory. Here $M$ is the ``geometrical'' mass of the spherical
gravitating body expressed in the convenient ``geometrical'' units $G = 1 = c$. It is
connected to the mass ${\cal M}$ expressed in the standard units by $M = G{\cal M}/c^2$ and
has the dimension of length.
\vskip0.2truecm \noindent
Another point discussed in this paper is based on the following two remarks: ({\it
i})\,Obviously, Newton's gravity theory is a limit of Einstein's general theory of
relativity. Should the limit {\it necessarily} correspond to ${\cal G} = 0$? Perhaps not,
because Newtonian physicists could discover {\it within} Newton's theory that ${\cal G} \not
= 0$. ({\it ii})\,They could also discover that the curvature of space depends on the
distance from the gravity center, ${\cal G} = {\cal G}(r)$. This would suggest to them,
again {\it within} the framework of Newton's theory, that gravity and curvature are not
independent, but instead they are somehow linked. Here we suggest that it is possible to
establish the link within an ``enhanced'' version of Newton's theory, by adding to its
standard version defined by (\ref{Poisson-equation}) and (\ref{equation-of-motion}) an extra
equation,
\begin{equation}
{}^{3}R = 2 k \rho ,
\label{EGR-Ricci}
\end{equation}
where ${}^{3}R $ is the Ricci scalar corresponding to ${}^{3}g_{ik}$, $\rho$ is the density
of matter, and $k$ is a constant. Equations (\ref{Poisson-equation}),
(\ref{equation-of-motion}) and (\ref{EGR-Ricci}) define our enhanced version of Newtonian
gravitational theory. In the special case of a spherically symmetric, vacuum ($\rho = 0$)
space, they {\it uniquely} lead to the 3-D metric of the form,
\begin{equation}
ds^2 = \left( 1 - \frac{r_0}{r}\right)^{-1}dr^2 + r^2 \left( d\theta^2 + \sin^2 \theta
d\phi^2\right) ,
\label{our-metric}
\end{equation}
where $r_0$ is a constant. A choice $r_0 = 4\,M$ leads to correct values for both the
perihelion advance and the light bending effects\footnote{Assuming that light moves along
geodesic lines in space.}.
%
\section{The three radii of a circle}
%
%
\begin{figure}[t!]
\begin{center}
\includegraphics[width=0.8\textwidth]{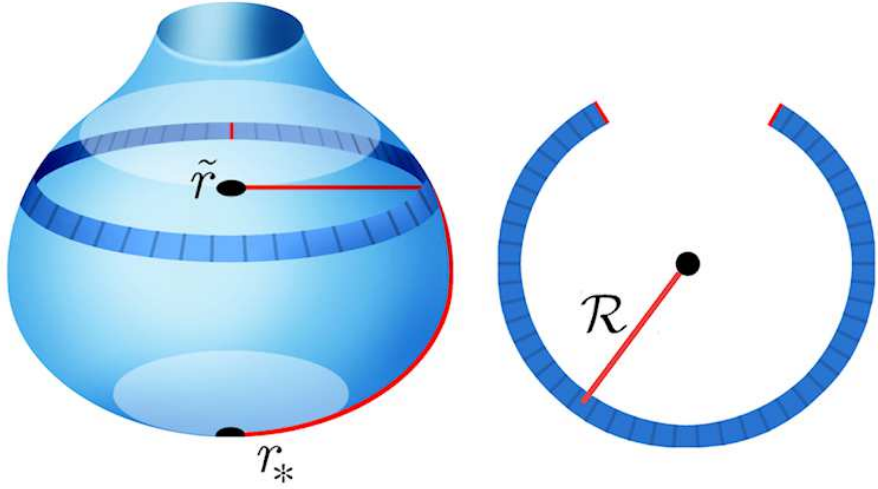}
\caption{For a circle placed in a curved space (here on a curved 2-D surface), its geodesic
radius $r_*$, circumferential radius ${\tilde r}$, and curvature radius ${\cal R}$ are all
different, $r_* \not = {\tilde r} \not = {\cal R}$. }
\end{center}
\label{Figure-three-radii}
\end{figure}
Consider a two dimensional curved, axisymmetric surface with the metric
\begin{equation}
ds^2 = dr_*^2 + [{\tilde r(r_*)}]^2d\phi^2.
\label{metric}
\end{equation}
and a family of concentric circles $r_* =$\,const in it. One of them is shown in
Figure~\ref{Figure-three-radii}. Obviously, $r_*$ is the geodesic radius and ${\tilde r}$ is
the circumferential radius of these circles,
\begin{eqnarray}
({\rm geodesic~radius}) &\equiv& \int_0^{r_*} ds_{\vert \phi = {\rm const}} = r_*,
\label{definition-r-star} \\
({\rm circumferential~radius}) &\equiv& \frac{1}{2\pi}\int_0^{2\pi} ds_{\vert r_* = {\rm
const}} = {\tilde r}.
\label{definition-r-tilde}
\end{eqnarray}
Let $\tau^i = {\tilde r}^{-1}\delta^i_{\,\phi}$ be a unit vector tangent to the circle. From
the Frenet formula,
\begin{equation}
\frac{d\tau^i}{ds} = - \frac{1}{{\cal R}} \lambda^i, ~{\rm where}~\lambda^i = ({\rm
unit~normal~to~the~circle}),
\label{curvature-radius}
\end{equation}
one deduces that the curvature radius ${\cal R}$ may be defined by,
\begin{equation}
({\rm curvature~radius}) \equiv \left[ \frac{d\tau^i}{ds}\,\frac{d\tau_i}{ds}\right]^{-1/2}
= {\cal R} .
\label{curvature-radius}
\end{equation}
Two useful formulae for the curvature of the circle, ${\cal K} = 1/{\cal R}$, and for the
Gaussian curvature ${\cal G}$ of the surface with the metric (\ref{metric}) read,
\begin{eqnarray}
%
{\cal K} = +\, \frac{1}{{\tilde r}}\left( \frac{d{\tilde r}}{\,dr_*} \right),
\label{radius-curvature-useful} \\
{\cal G} = -\, \frac{1}{{\tilde r}} \left( \frac{d^{2}{\tilde r}}{d r_*^2}\right).
\label{Gauss-curvarure-useful}
\end{eqnarray}
%
Formula (\ref{radius-curvature-useful}) follows from (\ref{curvature-radius}). For
derivation of (\ref{Gauss-curvarure-useful}) see e.g. \cite{Synge-1978}, Section 3.4.
\section{Equations of motion}
%
Let us consider a curve in space given by a parametric equation,
\begin{equation}
x^i = x^i(s),
\label{parametric-equation}
\end{equation}
where $x^i$ are coordinates in space, and $s$ is the length along the curve. If a body moves
along this curve, its velocity equals,
\begin{equation}
v^i = \frac{dx^i}{dt} = \frac{ds}{dt} \frac{dx^i}{ds} = v \tau^i.
\label{velocity}
\end{equation}
Here $v =ds/dt$ is the speed of the body and $\tau^i = dx^i/ds$ is a unit vector tangent to
the curve (\ref{parametric-equation}), i.e. the direction of motion. The acceleration may be
calculated as follow,
\begin{equation}
a^i = \frac{dv^i}{dt} = \frac{ds}{dt} \frac{d(v \tau^i)}{ds} = v^2 \left(
\frac{d\tau^i}{ds}\right) + \tau^i v\frac{dv}{ds}.
\label{acceleration}
\end{equation}
Assuming circular motion with constant velocity, $v = $ const, and applying
(\ref{curvature-radius}) to calculate the term in brackets, we arrive at
\begin{equation}
a^i = v^2\frac{1}{{\cal R}} \lambda^i,
\label{acceleration}
\end{equation}
which is the well known formula for the centrifugal acceleration.

Consider now circular motion around a spherically symmetric center of gravity. The Newtonian
equation of motion, $F^i = m a^i$, takes the form,
\begin{equation}
- \nabla^i \Phi = v^2\frac{1}{{\cal R}} \lambda^i,
\label{equation-motion}
\end{equation}
where $F^i = - m \nabla^i \Phi$ is the gravitational force, and $\Phi$ is the gravitational
potential. Three quantities characterize motion on a particular circular orbit: the angular
velocity $\Omega$, the angular speed $v$, and the specific angular momentum ${\cal L}$. They
are related by,
\begin{eqnarray}
%
%
v = {\tilde r} \Omega,
\label{velocity-speed-momentum-1}
\\
{\cal L} = {\tilde r} v = {\tilde r}^2 \Omega.
\label{velocity-speed-momentum-2}
\end{eqnarray}
Using (\ref{velocity-speed-momentum-2}), and multiplying its left side by $\lambda_i$, we
transform the equation of motion (\ref{equation-motion}) into a form which will be
convenient later,
\begin{equation}
\lambda_i \nabla^i \Phi = \frac{{\cal L}^2}{{\tilde r}^2 {\cal R}}.
\label{equation-motion-1}
\end{equation}
In this expression, $\lambda^i$ is a unit, {\it outside pointing}, vector. Here ``outside''
has the absolute meaning --- outside the center, in the direction towards infinity. We will
calculate the left-hand side of this equation in the next Section.

\section{Newton's gravity and Kepler's law}
In an empty space, the gravitational potential $\Phi$ obeys the Laplace equation,
\begin{equation}
\nabla_i (\nabla^i \Phi) = 0.
\label{Laplace}
\end{equation}
Let us integrate (\ref{Laplace}) over the volume $\mathbbm{V}$ that is contained between two
spheres, concentric with the gravity center, with sphere $\mathbbm{S}_1$ being inside sphere
$\mathbbm{S}_2$. We transform the volume integral into a surface integral, using the Gauss
theorem
\begin{equation}
0 = \int_{\mathbbm{V}}\nabla_i (\nabla^i \Phi) d\mathbbm{V}
= \int_{\mathbbm{S}_1}(\nabla^i \Phi) N^{(1)}_i d\mathbbm{S} +
\int_{\mathbbm{S}_2}(\nabla^i \Phi) N^{(2)}_i d\mathbbm{S}.
\label{Gauss}
\end{equation}
The oriented surface elements on the spherical surfaces $\mathbbm{S}_1$ and $\mathbbm{S}_2$
may be written, respectively, as
\begin{equation}
N^{(1)}_i d\mathbbm{S} = - \lambda_i d\mathbbm{S}, ~~N^{(2)}_i d\mathbbm{S} = + \lambda_i
d\mathbbm{S},
\label{surface-elements}
\end{equation}
therefore,
\begin{equation}
\int_{\mathbbm{S}_1}(\nabla^i \Phi) \lambda_i d\mathbbm{S} = \int_{\mathbbm{S}_2}(\nabla^i
\Phi) \lambda_i d\mathbbm{S}.
\label{two-spheres}
\end{equation}
This means that the value of the integral is the same, say $S_0$, for all spheres around the
gravity center. In addition, because of the spherical symmetry of the potential, the
quantity $(\nabla^i \Phi) \lambda_i$ is constant over the sphere of integration. Thus,
\begin{equation}
S_0 =  (\nabla^i \Phi) \lambda_i \int_{\mathbbm{S}} d\mathbbm{S} = 4\pi {\tilde r}^2\,
(\nabla^i \Phi) \lambda_i, ~~
(\nabla^i \Phi) \lambda_i = \frac{S_0}{4\pi {\tilde r}^2} = \frac{G{\cal M}}{{\tilde r}^2}.
\label{one-sphere}
\end{equation}
Combining (\ref{one-sphere}) with (\ref{equation-motion-1}), we may finally write,
\begin{equation}
{\cal L}^2 = G{\cal M}{\cal R}.
\label{Kepler-angular-momentum}
\end{equation}
This is the Kepler Third Law. Using natural units for radius and frequency,
\begin{equation}
R_G = \frac{G{\cal M}}{c^2} = M, ~~
\Omega_G = \frac{c^3}{G{\cal M}} = \frac{c}{M},
\label{gravitational-frequency}
\end{equation}
we may write the formula for the Keplerian angular velocity as,
\begin{equation}
\left(\frac{{\Omega}}{\,\Omega_G}\right)^2 = R_G^3 \left(\frac{{\cal R}}{{\tilde
r}^4}\right).
\label{Kepler-angular-velocity}
\end{equation}
%
\section{Epicyclic oscillations, the perihelion advance }
%
Suppose that we slightly perturb a test-body on a circular orbit. This means that its
angular momentum will not correspond to the Keplerian one, ${\cal L}^2$, given by
(\ref{Kepler-angular-momentum}), but will be slightly different ${\cal L}^2 + \delta{\cal
L}^2$. There will be also a small radial motion with velocity ${\dot{(\delta r_*)}}$ and
acceleration ${\ddot{(\delta r_*)}}$. From (\ref{equation-motion-1}) it follows that
\begin{equation}
\frac{GM}{{\tilde r}^2} - \frac{{\cal L}^2 + \delta{\cal L}^2}{{\tilde r}^2 {\cal R}} =
{\ddot{(\delta r_*)}}.
\label{perturbation}
\end{equation}
Keeping the first order term in equation (\ref{perturbation}), and using
\begin{equation}
\delta{\cal L}^2 = \frac{d{\cal L}^2}{d r_*}\,(\delta r_*),
\label{perturbation-delta-L}
\end{equation}
we arrive at the simple harmonic oscillator equation,
\begin{equation}
\omega^2 (\delta r_*) + {\ddot{(\delta r_*)}} = 0,
\label{simple-harmonic}
\end{equation}
where $\omega$ is the radial epicyclic frequency,
\begin{equation}
\omega^2 = \frac{1}{{\tilde r}^2 {\cal R}}\left(\frac{d{\cal L}^2}{d r_*}\right).
\label{epicyclic-frequency-1}
\end{equation}
Using equations (\ref{Kepler-angular-momentum}) and (\ref{gravitational-frequency}), we may
write the expression for the epicyclic frequency in the form,
\begin{equation}
\left( \frac{\omega}{\Omega_G}\right)^2 = \left(\frac{d{\cal R}}{\,\,d r_*}\right) \frac
{R_G^3}{{\tilde r}^2 \,{\cal R}},
\label{epicyclic-frequency-2}
\end{equation}
or comparing this with (\ref{Kepler-angular-velocity}),
\begin{equation}
\left( \frac{\omega}{\Omega}\right)^2 \,=\, \left(\frac{d{\cal R}}{\,\,d r_*}\right) \frac
{{\tilde r}^2}{{\cal R}^2} \,=\, \left(\frac{d {\tilde r}}{\,\,d r_*} \right)^2 - {\tilde r}
\left( \frac{d^2 {\tilde r}}{d r_*^2} \right).
\label{epicyclic-frequency-3}
\end{equation}
In a flat space, $r_*$$\,=\,$${\tilde r}$$\,=\,$${\cal R}$, and therefore
$\omega$$\,=\,$$\Omega$, which implies that the slightly non-circular orbit is a closed
curve, indeed an ellipse. In a curved space with ${\cal G}$$\,\not=\,$$0$, one has
$r_*$$\,\not=\,$${\tilde r}$$\,\not=\,$${\cal R}$, and consequently
$\omega$$\,\not=\,$$\Omega$. The slightly non-circular orbit would not be a closed curve. It
could be represented by a precessing ellipse, with two consecutive perihelia shifted by
\begin{equation}
\Delta \phi = T(\Omega - \omega) = 2\pi(1 - \frac{\omega}{\Omega})
= 2\pi \left[ 1 - \left( \frac{ d{\cal R} }{d{\tilde r} } \frac{\tilde{r} ^3}{\mathcal{
R}^3} \right)^{1/2} \right] ,
\label{perihelion-shift-1}
\end{equation}
where $T = 2\pi/\Omega$ is the orbital period.
%
\section{A Newtonian experiment}
\label{Newtonian-experiment}
%
Newtonian dynamics allows one to {\it measure} the circumferential ${\tilde r}$ and
curvature ${\cal R}$ radii of circular orbits by measuring the gravitational $a_G$ and
centrifugal $a_C$ radial accelerations for a circular orbit,
\begin{equation}
a_G = -\frac{G{\cal M}}{{\tilde r}^2}, ~~ a_C = \frac{V^2}{{\cal R}}.
\label{gravitational-centrifugal}
\end{equation}
In the Schwarzschild metric, the acceleration of a particle (a ``planet'') moving with the
orbital velocity $v$ along a circular orbit equals,
\begin{equation}
a_i = \nabla_i \Psi + V^2 \frac{\nabla_i \mathbf{R}}{\mathbf{R}}.
\label{Schwarzschild-acceleration}
\end{equation}
Here $V = v/(1-v^2)^{1/2}$, and the scalars $\Psi$ and $\mathbf{R}$ are expressed in terms
of the time-symmetry Killing vector $\eta^i$, and the axial-symmetry Killing vector $\xi^i$,
\begin{equation}
\Psi = -\frac{1}{2}\ln (\eta^i\,\eta_i), ~~ \mathbf{R}^2 = -
\frac{(\xi^k\,\xi_k)}{(\eta^i\,\eta_i)}.
\label{Schwarzschild-definitions}
\end{equation}
In Schwarzschild coordinates this is, at the ``equatorial plane'' $\theta = \pi/2$,
\begin{equation}
(\eta^i\,\eta_i) = g_{tt} = 1 - \frac{2M}{r}, ~~  ~~ (\xi^k\,\xi_k) = - r^2.
\label{Schwarzschild-in-coordinates}
\end{equation}
This allows one to calculate the results of the Newtonian experiment to measure the
gravitational and centrifugal accelerations,
\begin{equation}
a_G = -\frac{1}{2}\frac{d}{dr}\left[\ln \left( 1 - 2M/r\right)\right], ~~
a_C = \frac{1}{2}\, V^2\, \frac{d}{dr}\left[\ln \left( \frac{r^2}{1 - 2M/r}\right)\right].
\label{Newtonian-results}
\end{equation}
By comparing (\ref{gravitational-centrifugal}) and  (\ref{Newtonian-results}), one concludes
that,
\begin{equation}
\tilde{r}(r) = r \left( 1-2M/r \right)^{1/2}, ~~
{\cal R}(r) = r \frac{ 1 - 2M/r}{1 - 3M/r}.
\label{Newtonian-results-two-radii}
\end{equation}
Note, that to ${\cal O}^1(M/r)$ accuracy this is $r = {\tilde r} = {\cal R}$. Therefore,
curvature effects may appear at this order.
We can also usefully calculate the derivative $d r_* (r) / dr$, as only the derivative, not
the absolute value of $r_* (r)$ is of interest. The following equation follows from the
definition of Frenet's curvature radius $\mathcal{R}$
\begin{equation}
\frac{dr_*}{dr} = \frac{\mathcal{R}}{\tilde{r}}\frac{d \tilde{r}}{d r} = \frac{r-M}{r-3M} .
 \label{NMgrr}
\end{equation}
The above formula allows one to write the metric of the 2-D space geometry, $ds^2 = dr^2_* +
{\tilde r}^2 d\phi^2$, measured in this Newtonian experiment,
\begin{equation}
ds^2 = \left(\frac{r-M}{r-3M} \right)^2 dr^2 + r^2\left(1-\frac{2M}{r}\right) d \phi^2 .
 \label{NewtMetric}
\end{equation}
Inserting (\ref{Newtonian-results-two-radii}) into the Newtonian perihelion advance formula
(\ref{perihelion-shift-1}) one gets,
\begin{equation}
\frac{\Delta \phi}{2 \pi} = 1 - \sqrt{1 + \frac{-6Mr^3 + 34M^2r^2 -62M^3r + 36M^4}{r^4 -
5Mr^3 + 8M^2r^2 -4M^3r}} .  \nonumber
\label{perAdvance}
\end{equation}
Expanding this to the desired accuracy ${\cal O}^2(M/r)$, one finally gets the same value
for the perihelion advance as calculated in Einstein's theory,
\begin{equation}
\Delta \phi = 6 \pi \frac{M}{r} + {\cal O}^2 \left(\frac{M}{r} \right).
\label{final-perihelion-maciek}
\end{equation}

\section{Light bending}
\label{light-bending}
Knowing the space geometry, given by equation (\ref{NewtMetric}), we may calculate the
effect of {\it light bending} assuming that light travels along geodesic lines in space. In
Newton's theory this assumption is equivalent to the {\it Fermat principle}, i.e. that light
travels (with a constant speed) between two points $A$, $B$ in space, minimizing the time
travel $T_{AB}$. The equation of motion for the $\phi$ coordinate is, in these
circumstances,
\begin{equation}
\frac{d^2 \phi}{ds^2} + 2 \frac{r-M}{r(r-2M)} \frac{dr}{ds} \frac{d\phi}{ds} = 0 ,
\end{equation}
from which we find $\frac{d \phi}{ds}$ to be equal to
\begin{equation}
\frac{d\phi}{ds} = \frac{C}{r(r-2M)} .
\end{equation}
The integration constant can be evaluated at the perihelion location $r = R_0$ (i.e. where
${d \phi}/{ds} = 0$), yielding
\begin{equation}
\frac{d\phi}{ds} = \frac{\sqrt{R_0(R_0-2M)}}{r(r-2M)} . \label{eq:phi}
\end{equation}
Using equations (\ref{NewtMetric}) and (\ref{eq:phi}) we find also
\begin{equation}
\frac{dr}{ds} = \frac{r-3M}{r-M}\sqrt{1 - \frac{R_0(R_0-2M)}{r(r-2M)}}.
\label{eq:r}
\end{equation}
After dividing equation (\ref{eq:phi}) by equation (\ref{eq:r}) and substituting $x = R_0/r$
the $d \phi / d r$ equation can be integrated from $R_0$ to $\infty$ (or $x$ from 0 to 1),
which will give us the half of $\pi + \delta$.
Let us also define $\mu = M/R_0$, then
\begin{equation}
\frac{\pi+ \delta}{2}
= \int^1_0 \frac{1 - x \mu}{(1 - 2 x \mu)(1- 3 x \mu)}
\sqrt{\frac{1 -2\mu}{1 - x^2(1-2\mu)/(1-2x \mu) }} d x .
\end{equation}
%
This integration can be expanded in a Taylor series for $\mu$:
\begin{equation}
\frac{\pi+ \delta}{2}
= \int^1_0 \frac{dx}{\sqrt{1-x^2}} + \int^1_0 \frac{(3x^2 +3x-1)dx}{\sqrt{1-x^2}(x+1)} \mu +
{\cal O}^2(\mu) .
\end{equation}
As the first component on the right hand side is equal to $\pi/2$, we conclude that
\begin{equation}
\delta \approx 2\mu \int^1_0 \frac{(3x^2+3x-1)dx}{\sqrt{1-x^2}(x+1)} = 4 \frac{M}{R_0} .
\end{equation}
Similar calculations in the Schwarzschild spacetime geometry give the same
result\footnote{Which is twice the well-known {\it flat-space} and massive photon Newtonian
prediction.}
\begin{equation}
\delta \approx 2\mu \int^1_0 \frac{1-x^3}{(1-x^2)^{3/2}} = 4 \frac{M}{R_0} .
\end{equation}
Once again, the prediction of the Newtonian theory in the non-flat space is found to be
consistent with observations (and with Einstein's general relativity).
\section{Enhanced Newtonian Gravitational Theory}
J{\"u}rgen Ehlers pointed out in 1961 that in Einstein's theory the curvature of the
rest-space of irrotational matter is determined by its distribution and relative motion (see
1221 in his article \cite{Ehlers93}). The equations governing such 3-space curvature for
arbitrary irrotational flows are given in \cite{EllVan98}; see their equation (54).
Consequently it makes sense to consider gravitational dynamics in the context of
3-dimensional curved Riemannian spaces. As Newtonian theory is an approximation to General
Relativity Theory, it is therefore interesting to see what happens in the case of Newtonian
theory in a curved 3-dimensional background space.

In the case of isometric flows, $\theta = \sigma_{ab} = 0$ and there is a potential such
that $\dot{u}^a = U_{,a}$ where the gravitational potential $U$ relates the Killing vector
$\xi$ to the unit 4-velocity $u^a$ by $\xi^a = e^U u^a$ (see 1234 in Ehlers
\cite{Ehlers93}). Then the relevant equation becomes
\begin{equation}
{}^3R_{ab} =  \tilde{\nabla}_a\tilde{\nabla}_b U \,+ \,\tilde{\nabla}_a U \tilde{\nabla} _b
U \,+ \,\frac{2}{3} \,k\,\rho \,h_{ab} ,
\label{3_curv}
\end{equation}
where $\tilde{\nabla}_a$ is the 3-dimensional covariant derivative,  $\rho$ is the energy
density of matter, and we have assumed anisotropic stress is zero ($\pi_{ab} = 0$) and a
vanishing cosmological constant. Here $h_{ab} = g_{ab} - u_a u_b$ is the metric of the
three-spaces orthogonal to $u^a$. This case will include static spherically symmetric
spacetimes. Taking the trace of this equation gives (see equation (55) in \cite{EllVan98})
\begin{equation}
{}^3R  =  2\, k \,\rho ,
\label{3_curv_salar}
\end{equation}
where the potential terms have gone because of the relation between the 3-dimensional and
4-dimensional covariant derivatives. Together with the Poisson equation and equation of
motion it defines the \emph{Enhanced Newtonian Gravitational Theory},
\begin{eqnarray}
%
%
{}^{3}g^{ik}\,\, \nabla_i \nabla_k \Phi = -4\pi G \rho,
%
\nonumber \\
F_i = m a_i, \nonumber \\
{}^3R  =  2 k \,\rho .
\label{enhanced}
\end{eqnarray}
For spherically symmetric spaces, the most general metric has the form,
\begin{equation}
ds^2 = A(r)\,dr^2 + r^2 \left( d\theta^2 + \sin^2 \theta d\phi^2 \right),
\label{A-metric}
\end{equation}
and in the vacuum case, $\rho = 0 = {}^3R$ one has
\begin{eqnarray}
\hskip0.2truecm
{}^3R_{rr} &=& \frac{A^\prime}{rA}, \label{Ricci-components-01} \\
{}^3R_{\theta \theta} &=& \left( \frac{rA^\prime}{2A^2} -
\frac{1}{A} + 1 \right)\sin^2\theta, \label{Ricci-components-02} \\
{}^3R_{\phi \phi} &=&  \frac{rA^\prime}{2A^2} - \frac{1}{A} + 1,
\label{Ricci-components-04} \\
{}^3R &=& \frac{2}{r^2} - \frac{2}{A r^2} + \frac{2A^\prime}{A^2 r} = 0.
\label{Ricci-components-04}
\end{eqnarray}
Here the prime denotes a derivative with respect to $r$. Equation
(\ref{Ricci-components-04}) has a unique solution,
\begin{equation}
A(r) = \left( 1 - \frac{r_0}{r} \right)^{-1},
\label{A-solution}
\end{equation}
with $r_0$ being an integration constant. Its value cannot be determined by equations
(\ref{enhanced}), but instead must be chosen by correspondence with experiment\footnote{In
Einstein's theory, when one derives the Schwarzschild metric, a constant of integration is
determined in a similar way, i.e by correspondence with Newton's theory.}. Using the same
procedure as in Sections \ref{Newtonian-experiment} and \ref{light-bending}, one proves that
the choice $r_0 = 4M$ gives the correct values for the perihelion advance and light bending
(with accuracy ${\cal O}(r_0/r)$).
%
\section{The two metrics}
We have shown that ``experimentally'' established and the ``theoretically'' postulated
Newtonian metrics of the curved 3-D space corresponding to a spherically symmetric body are,
respectively,
\begin{eqnarray}
\hskip0.0truecm
ds^2 &=& \left(\frac{r-M}{r-3M} \right)^2 dr^2 + r^2\left(1-\frac{2M}{r}\right)\left(
d\theta^2 + \sin^2 \theta d\phi^2\right),
\nonumber \\
&~& \label{metric-experimental} \\
ds^2 &=& \left( 1 - \frac{4 M}{r}\right)^{-1}dr^2 + r^2 \left( d\theta^2 + \sin^2 \theta
d\phi^2\right).
\label{metric-theoretical}
\end{eqnarray}
We have also shown that {\it any} spherically symmetric metric that obeys $^3 R = 0$ must be
isometric with (\ref{metric-theoretical}). The Ricci scalar for the ``experimental'' metric
may be calculated to be
\begin{eqnarray}
({}^3 R)M^2 &=& M^2\frac{18(M/r)^3 - 10(M/r)^2}{r^2[ 4(M/r)^3 - 8(M/r)^2 +5(M/r) - 1]}
\nonumber \\
&=& 10(M/r)^4 + 32(M/r)^5 + ...  \nonumber \\
&=& 0 + {\cal O}^4(M/r).
\label{experimental-Ricci}
\end{eqnarray}
On the other hand a metric,
\begin{equation}
\hskip0.0truecm
ds^2 =\frac{dr^2}{1 - \frac{4 M}{r} + \alpha \left(\frac{M}{r}\right)^2
+ \beta \left(\frac{M}{r}\right)^3 + ...}
+ r^2 \left( d\theta^2 + \sin^2 \theta d\phi^2\right),
\label{metric-theoretical-A-B}
\end{equation}
has the Ricci tensor,
\begin{equation}
({}^3 R)M^2 = 2\,\alpha \left( \frac{M}{r}\right)^4
+ 4\,\beta \left( \frac{M}{r}\right)^5 + ...
= 0 + {\cal O}^4(M/r).
\label{Ricci-A-B}
\end{equation}
Thus, the experimental metric (\ref{metric-experimental}) and the theoretical metric
(\ref{metric-theoretical}) describe, with accuracy ${\cal O}^2(M/r)$, {\it the same}
geometry of space.
\section{Conclusions}
We demonstrated that a Newtonian physicist may experimentally determine the geometry of the
3-D space $^3g_{ik}^E$ by measuring gravitational and centrifugal accelerations. He may then
predict by calculations  the perihelion advance and the light bending as effects of the
curvature of space. The predicted values agree with the ones measured. We also demonstrated
that one may extend Newton's theory of gravitation by adding an equation that links Ricci
curvature of space with the density of matter. We calculated the resulting theoretical
metric of space $^3g_{ik}^T$ assuming spherical symmetry. In this metric, the values of
perihelion advance and light bending also agree with those observed. The two metrics
represent the same geometry, $^3g_{ik}^E = {}^3g_{ik}^T$ with accuracy ${\cal O}^2(M/r)$.
\vskip 0.2truecm \noindent
Abramowicz \cite{Abramowicz-2012} has shown that for spaces with constant Gaussian curvature
Newton's theory predicts no perihelion advance. We speculate that this is why Gauss (and
other XIX century mathematicians) who might have calculated Newtonian orbits in curved
spaces, would have missed the effect of perihelion advance. Most probably, they would
calculate orbits in spaces with a constant Gaussian curvature first. Gauss almost certainly
made this calculation. He was a master in calculating orbits. He made himself famous at the
age of 23 by calculating the orbit of Ceres, discovered in 1801 by Piazzi. He seriously
considered the possibility that our space is curved. He even attempted to determine the
curvature of space by measuring angles in a big triangle (69 km, 84 km, 106 km) made by the
summits of Brocken, Hoher Hagen and Gro{\ss}er Inselsberg. Gauss was not quick in publishing
his results concerning curved spaces. It is known that he discovered most of Bolyai's
results, but never published them. Gauss died in February 1885, four years before Le Verrier
discovered the effect of the perihelion of Mercury advance.
%
\ack
The work presented here was started at the Gastroenterology and Trans\-plan\-tology Ward of
the MSW Hospital in Warsaw, before and after MAA's surgery. It was continued at the
Institute of Astronomy in Prague and finished at the University of Cape Town, where the work
was supported by the South African National Research Foundation (NRF) and the University of
Cape Town. The work has been supported by the Polish National Health Foundation (NFZ) and
the NCN UMO-2011/01/B/ST9/05439 Polish grant. Work of MW was partially
supported by the European Union in the framework of European Social Fund through the Warsaw
University of Technology Development Programme.
MAA thanks Dr Andrzej Otto who performed the surgery at the MSW Hospital in Warsaw.
%
\section*{References}


\end{document}